\documentstyle[11pt,emulateapj]{article}

\def\msun{{\,M_\odot}}
\def\cm{{\rm \,cm}}
\def\ghz{{\rm \,GHz}}
\def\mhz{{\rm \,MHz}}
\def\K{{\rm \,K}}
\def\sec{{\rm \,s}}
\def\erg{{\rm \,erg}}
\def\kev{{\rm \,keV}}
\def\mjy{{\rm \,mJy}}
\def\gm{{\rm \,g}}
\def\yr{{\rm \,yr}}
\def\dm{{\dot{M}}}
\def\mpc{{\rm \,Mpc}}
\def\pc{{\rm \,pc}}
\def\kpc{{\rm \,kpc}}
\def\kms{{\rm \,km\,sec}^{-1}}
\def\ngc{{ NGC 4258 }}
\def\eps{{\epsilon}}
\def\<{{\langle}}
\def\>{{\rangle}}

\def\M{{G M/c^2}}
\def\am1{{\alpha_{-1}}}
\def\ep1{{\epsilon_{-1}}}

\begin{document}

\title{What is the Accretion Rate in NGC 4258?}

\author{Charles F. Gammie \altaffilmark{1} and Ramesh Narayan}
\affil{Harvard-Smithsonian Center for Astrophysics, MS-51 \\
60 Garden St., Cambridge, MA 02138}
\and
\author{Roger Blandford}
\affil{Theoretical Astrophysics, Caltech 130-33, Pasadena, CA 91125}

\altaffiltext{1}{Also Isaac Newton Institute, 20 Clarkson Rd.,
Cambridge CB3 0EH, UK}

\begin{abstract}

We consider the implications of recent infrared and radio observations
of the nucleus of NGC~4258.  There is no direct evidence that the nucleus
has been steadily accreting on the viscous timescale of the outer masing
disk, which is $\gtrsim 10^9 \yr$.  Thus the mass accretion rate in the
outer disk need not be the same as in the inner accretion flow where
most of the gravitational binding energy is released.  We show that an
advection-dominated flow model with a transition radius of $\sim (10-100)
\M$ (where $M$ is the mass of the hole) and $\dm \approx 10^{-2}\msun
\yr^{-1}$ is consistent with the observed spectrum from radio to X-rays.
We also show that a thin (flat or warped) disk can fit the observed
fluxes outside the X-ray band.  The X-rays can be explained by means
of a corona in such a model, but the absence of radio emission from
the location of the putative central black hole provides a serious
constraint on the properties of the corona. A wide range of accretion
rates, $10^{-4} \lesssim \dm \lesssim 10^{-2} \msun\yr^{-1}$, can be
made to fit the data, but the most ``natural'' models have $10^{-3}
\lesssim \dm \lesssim 10^{-2} \msun\yr^{-1}$.  Physical conditions in
the observed VLBI jet features can also be related to conditions in the
inner accretion flow. We conclude with a list of future observations
that might help to constrain the accretion rate.

\end{abstract}


\section{Introduction}

Most measurements of the physical properties of astronomical objects
are fortunate to obtain an accuracy of a factor of two.  The rare
instances when more precise measurements are possible offer significant
opportunities for advancing our knowledge, and so it is worth investing
some effort to understand these objects in detail.  The orbiting water
maser spots in \ngc provide one of these rare instances.

Water maser emission was first detected in the nucleus of \ngc by
\cite{chl84}.  It was soon realized that the masers might lie in a disk
(\cite{cl86}).  \cite{nim93} discovered high velocity maser emission,
which \cite{ww94} quickly interpreted as a natural outcome of maser
emission from a disk.  VLBI observations (\cite{gr95b}, \cite{mi95},
\cite{mo95}, \cite{hgm96}) demonstrated that the maser spots are
positioned on the sky as one would expect for a nearly edge-on thin disk.
A model fit to the maser spots allows one to determine the central mass
(\cite{h97}).  By measuring maser spot accelerations (\cite{hb90},
\cite{ha94}, \cite{gr95c}, \cite{gr95a}, \cite{h97}), or proper motions
(\cite{h97}) one can also directly determine a geometric distance.

The measurement of an accurate mass and distance eliminate two of the
principal uncertainties that plague efforts to understand the physics
of active galactic nuclei (AGN).  The mass of the central object in
\ngc (which we will freely refer to as a black hole) is $(4 \pm 0.25)
\times 10^7 \msun$ (\cite{h97}; $1\sigma$).  The distance is measured
to be $7.3 \pm 0.3 \mpc$, with independent and consistent results from
measurements of maser spot acceleration and maser spot proper motion
(\cite{hp97}).

One of the other main uncertainties that plagues AGN studies is 
the accretion rate $\dm$.  The accretion rate in \ngc has been
controversial, and there are a number of estimates in the literature:
\cite{nm95}(hereafter NM), $7 \times 10^{-6} \alpha_{-1} \msun \yr^{-1}$, where
$\alpha = 0.1 \am1 \lesssim 1$ is the usual dimensionless angular momentum
diffusion coefficient of accretion disk theory (\cite{ss73}, \cite{jep81});
\cite{la96}, $0.014 \am1 \msun \yr^{-1}$ ;
\cite{cj97}, $8 \times 10^{-5} \am1 \msun \yr^{-1}$ ;
\cite{mm97}, $7 \times 10^{-3} \msun \yr^{-1}$ ;
\cite{pk97}, $1.5 \times 10^{-3} \msun \yr^{-1}$.
These estimates differ because of disagreements about the nature
and efficiency of angular momentum transport in disks and, to a
lesser degree, because of disagreements about the heating and cooling
physics in various parts of the accretion flow.  The purpose of this
paper is to clarify some of these differences, propose observational
tests that reduce the number of acceptable models, and thereby narrow
the range of uncertainty in $\dm$.

We begin in \S2 by asking what the masing disk might tell us about the
accretion rate.  In \S 3 we give a critical review of the pertinent
data;  in \S 4 we consider models for the inner accretion flow; in \S 5
we consider implications of observations of the jet.  A discussion and
summary of interesting observational and theoretical problems follows
in \S 6.

\section{Masing Disk}

Most estimates of the accretion rate in \ngc are based on observations
of the masing disk and the assumption that the disk is in a steady
state ($d \dm(r)/dr = 0$) from the masing disk in to the black hole,
where most of the gravitational binding energy is released (NM,
\cite{cj97,mm97,pk97}).  But is the masing disk in a steady state?  In
the absence of significant external torques (e.g. a magnetohydrodynamic
wind) one expects a steady state to be achieved on the viscous
timescale $t_v \equiv (\alpha\Omega)^{-1} \times (r/H)^2$, if mass is
steadily supplied to the disk from $r \gtrsim 0.26 \pc$.

In the masing disk, $t_v = 2.6 \times 10^{9} \am1^{-1} (200 \K/T) (r/0.2
\pc)^{1/2} \yr$.  \footnote{An optically thin coronal component with $T
\sim 8000 \K$, as suggested by NM, might carry a large $\dm$ with a small
viscous timescale.  But in the NM model, at least, most of the mass flux
is carried in the cool, optically thick component at the inner edge of the
masing disk.} This is a significant fraction of a Hubble time.  We know
of no direct evidence that mass has been steadily supplied to the nucleus
of \ngc, or to any active nucleus, on this timescale.  Thus there is no
direct support for the assumption that the disk is in a steady state.

\ngc contains an unusual, $\kpc$-scale jet detected in optical lines
(\cite{cc61}), radio continuum (\cite{kom72}), and X-rays (\cite{cwd95}).
If the jet is produced in the immediate vicinity of the black hole, and
the jet mass loss rate is closely coupled to the accretion rate, as is
commonly believed, then it provides a fossil record of past accretion.
The characteristic timescale for the jet is $t_J = R/V_J = 2.4 \times 10^6
(R/5\kpc)(2000 \kms/V_J) \yr$ where $V_J$ is the jet radial velocity.
More detailed models of the jet suggest an age of at least $10^6$ to
$10^7\yr$ (e.g.  \cite{vv82,mrnl89}).  Thus the jet provides a record over
only a small fraction of the masing disk viscous timescale.  The jet is
roughly continuous over its length, however (there are no lengthy gaps
in the jets: see for example Figure 2 of \cite{cwd95}), which suggests
approximately steady accretion over a time $t_J$.  For standard $\alpha$
disk models for the inner accretion flow we find $t_v = t_J \sim 10^6
\yr$ at $r \sim 10^3 G M/c^2$ (the precise value depends on the accretion
rate, $\alpha$, etc.).  Thus only a small region of the accretion flow
in \ngc needs to be in a steady state.

\section{Critical Review of Data}

\subsection{Nuclear Spectral Energy Distribution}

\cite{cb97} have recently reported J,H, and K band ($1.25, 1.65, 2.21
\micron$) observations of \ngc in $0.6''$ seeing.  Using the J band
image as a baseline, they subtract a scaled J band image from the H and K
images and find a nuclear emission excess consistent with a point source.
The excess is $1.1 \mjy$ at H and $4.5 \mjy$ at K.  It is then possible
to deduce the extinction and intrinsic luminosity of the source if the
extinction follows a standard form deduced from the local interstellar
medium (\cite{ccm89}) and if the intrinsic spectrum has an approximately
power-law form over the region of interest ($f_\nu \sim \nu^s$):
\begin{equation}
A_V = 20.1 (1 + 0.21 s) {\rm mag},
\end{equation}
\begin{equation}
\log(\nu L_\nu(H)/ \erg \sec^{-1}) = 41.63 + 0.32 s,
\end{equation}
\begin{equation}
\log(\nu L_\nu(K)/ \erg \sec^{-1}) = 41.51 + 0.19 s.
\end{equation}
The spectral index $s$ might reasonably range from $\sim 1/3$ for the
classical thin disk spectrum to $\sim -1$, which is a typical optical
spectral index for an unobscured AGN.  The implied flux at J is $0.2
(2.62)^s \mjy$, which is small, justifying the use of the J band image as
a baseline.  Taking $R_V = 3.1$ (\cite{spitzer}), the extinction implies
an obscuring column
\begin{equation}
N_H = 3.8 \times 10^{22} (1 + 0.21 s) (Z_\odot/Z) \cm^{-2}
\end{equation}
($Z\equiv $ metallicity) or, assuming cosmic abundance of helium,
\begin{equation}
\Sigma = 0.09 (1 + 0.21 s) (Z_\odot/Z) \gm \cm^{-2}.
\end{equation}

These estimates are sensitive to the assumed extinction curve, since
\begin{equation}
{d\ln \nu L_\nu (H) \over{d \ln (A_H/A_K)}} = -5.3 - 1.1 s,
\end{equation}
and
\begin{equation}
{d A_V \over{d \ln (A_H/A_K)}} = -50 - 10 s.
\end{equation}
Evidently $10\%$ variations in extinction can lead to factor-of-two
errors in the intrinsic luminosity.

\cite{ma94} have reported X-ray observations of \ngc with ASCA.
They fit a four component model to the spectrum, including an obscured
power law, an iron line, a Raymond-Smith plasma, and thermal
bremsstrahlung.  They find an integrated luminosity for the power law
component in the $2$ to $10 \kev$ band of $(4.4 \pm 1.1) \times 10^{40}
\erg \sec^{-1}$ (at $7.3 \mpc$), and an absorbing column $N_H = (1.5
\pm 0.2) \times 10^{23} \cm^{-2} Z_\odot/Z$.  Assuming a cosmic
abundance of helium, this implies $\Sigma = 0.36 \gm \cm^{-2}
(Z_\odot/Z)$.  The X-ray column is a factor of $4$ larger than the column
inferred from infrared extinction which, given the uncertainties, may
be considered reasonably good agreement.  Indeed, the two columns need
not agree if the IR emitting material is also producing the X-ray
absorption.  The integrated luminosity implies $\nu L_\nu = 2.7 \times
10^{40} \erg \sec^{-1}$ at $4.5 \kev$, taking the mean of $\nu L_\nu$
across the $2$ to $10 \kev$ band.

Makishima et al.'s data are now publicly available through the ASCA
data archives.  We have tried fitting only an absorbed power law to the
$2.5$ to $10 \kev$ data.  While the total luminosity is robust, being
mostly determined by the flux near $10 \kev$, our absorption column is
lower than the Makishima et al. value.  This suggests that the derived
absorption column is sensitive to the presence of the thermal
bremsstrahlung component in the Makishima et al. model, and may be
somewhat less reliable than the total luminosity.

\cite{h98} report a $3 \sigma$ upper limit on the $22 \ghz$ flux from
the neighborhood of the compact object of $0.22 \mjy$, or $\nu L_\nu =
3.1 \times 10^{35} \erg \sec^{-1}$.

\cite{rl78} report a detection in a $5''$ aperture at $10 \micron$ of
$100 \mjy$, or $\nu L_\nu = 1.9 \times 10^{41} \erg \sec^{-1}$.  A
consistent result at similar aperture size is reported by
\cite{cwb85}.  Because of the large aperture these are upper limits.

Finally, the nucleus has been detected in polarized optical light by
\cite{w95}.  The polarized continuum has $f_\nu \sim \nu^{-1.1 \pm
0.2}$, with an amplitude $\nu L_\nu = 1.1 \times 10^{39} \erg
\sec^{-1}$.  The polarized spectral lines have FWHM of order $2500
\kms$, a velocity characteristic of radii just within the radius of
the masing disk.  These results offer only weak constraints on $\dm$
because the composition and distribution of the scattering medium are
poorly constrained.  Following Wilkes et al. we will regard the
polarized emission as providing an upper limit on $\nu L_\nu$ of $4
\times 10^{43} \erg \sec^{-1}$.

\subsection{Jet}

The jet in \ngc is observed at milliarcsecond scales in $22 \ghz$
continuum with the VLBA (\cite{he97}) and at arcsecond and larger scales
in the radio, optical, and X-ray.

The VLBA jet was discovered by \cite{he97}, while higher
signal-to-noise data has recently been presented by \cite{h98}.  The
northern jet has flux density $\simeq 3 \mjy$, is variable on a
timescale of weeks, and is unresolved.  Its mean location is about
$0.014 \pc$ ($ = 17$ light-days $= 7500 G M/c^2$) north of the best-fit
location for the compact object, and its position varies significantly
with time.  The southern jet has flux density $\simeq 0.5 \mjy$, has
not varied significantly in flux or position, and is also unresolved.
It is located about $0.035 \pc$ ($ = 43$ light-days $= 19,000 G M/c^2$)
south of the best-fit location for the compact object.  The difference
in brightness between northern and southern jet is attributed by
\cite{he97} to free-free absorption in the masing disk.  The emitting
blobs are approximately due south and north of the compact central
object (\cite{h98}).  This is consistent with the jet being normal to
the outer, masing disk and suggests that the disk is not strongly
warped, at least in position angle, between the masing disk and the
radius where the jet is launched.

Radio and X-ray observations show a curious, twisted jet outflow both
to the north and the south extending out to $\sim5$~kpc (summarized in
\cite{cwt92,cwd95,cmv95}).  The jet X-ray emission has a power
$\sim2\times10^{40}$~erg s$^{-1}$ and can be satisfactorily fit by a
thermal spectrum with temperature $T_X\sim3\times10^6$~K.  Associated
with this jet are forbidden emission lines which appear to come from
locally photoionized shocked plasma.  The line widths suggest
velocities $\sim500$~km s$^{-1}$.

\section{Inner Disk Models}

An observed flux $F_{obs}$ suggests a minimum accretion rate
$\dot{M}_{min} = 4\pi d^2 F_{obs}/(\eps c^2)$, where $\eps = 0.1 \ep1$ is the
accretion efficiency.  This is not a rigorous limit since the flow can
emit anisotropically and a spinning black hole can provide an additional
reservoir of energy.  Setting aside these concerns for the moment, it 
is reasonable to take $\ep1 \sim 1$ (recall
that the specific binding energy on the last stable circular orbit
varies from $0.038 c^2$ for retrograde orbits around a maximally
rotating black hole to $0.42 c^2$ for prograde orbits around a
maximally rotating hole; this binding energy is approximately related
to the efficiency $\eps$ of a thin disk accretion flow).
Taking the observed flux densities in the infrared,
integrating over the band, and adding this to the $2-10 \kev$ flux
(while this is corrected for absorption, the total flux is not
sensitive to the absorption), we find $\dot{M}_{min} = 10^{-5}
\ep1^{-1} \msun \yr^{-1}$.  This assumes the rest of the spectrum is
black (which is not correct given the detection of scattered light in
the optical by \cite{w95}) and the infrared extinction is zero.  A less
radically skeptical approach is to set the luminosity equal to the
extinction-corrected infrared $\nu L_\nu$.  Then $\dot{M}_{min} = 7
\times 10^{-5} (2.1)^s \ep1^{-1} \msun \yr^{-1}$.

The rest of the spectrum is not black.  By analogy with other AGN, \ngc
should have a rather flat spectrum in $\nu L_\nu$.  One can then ask what
is the minimum luminosity in a power law spectral energy distribution
that extends from the optical to $100 \kev$ and passes through the X-ray
data point;  we find the minimal power law slope is $\nu L_\nu \sim
\nu^{0.27}$, and the corresponding luminosity is $2.3 \times 10^{41} \erg
\sec^{-1}$.  Adding this to the minimal accretion rate from the infrared
point gives $\dot{M}_{min} = (7 (2.1)^s + 4) \times 10^{-5}\ep1^{-1}
\msun \yr^{-1},$ although some of the optical, UV, and soft X-ray flux
may be absorbed by a disk and reradiated in the infrared.

So far we have assumed that the accretion flow radiates isotropically,
but a thin disk, for example, has $f_\nu \sim \cos(i)$, if we ignore the
effects of limb-brightening.  Since the masing disk inclination is $82
^{\circ}$, and the most natural assumption is that the inner disk has the
same inclination (although there is no direct evidence for this), there
will also be a large additional upward correction in the mass accretion
rate, since $\cos(82 ^{\circ}) \simeq 0.14$.  A specific model is required
for a quantitative estimate of the inclination correction.

\subsection{Thin Disk Model}

Suppose that the central engine is just a thin, flat disk that radiates
locally like a black body.  On the $\nu^{1/3}$ portion of the classical
thin disk spectrum,
\begin{equation}\label{FNEQN}
f_\nu = 17.8 \dm_{-4}^{2/3} \cos(i) \lambda^{-1/3} \mjy,
\end{equation}
where $\lambda$ is in microns and $\dm_{-4}$ is the accretion rate in
units of $10^{-4}\msun\yr^{-1}$.  Inverting this equation, the required
mass accretion rate is
\begin{equation}
\dm \simeq 1.3 \times 10^{-6} f_\nu^{3/2} \lambda^{1/2}
	\cos(i)^{-3/2} \msun \yr^{-1},
\end{equation}
where $f_\nu$ is in $\mjy$.  Assuming $f_\nu \sim \nu^{1/3}$ and $i =
82^{\circ}$, the infrared data of Chary \& Becklin require $\dm = 0.011
\msun \yr^{-1}$.  Notice that the inclination correction is quite large
because $\dm$ depends nonlinearly on $\cos(i)$.  This also assumes the
disk extends down to $r = 0$; if the disk is truncated properly (say
at $r = 6 G M/c^2$) then $\dm$ must be evaluated numerically:  $\dm =
0.013 \msun \yr^{-1}$.  Thin disk spectra for $i = 0^\circ, 45^\circ,$ 
and $82^\circ$ are shown in Figure 1.

There are several effects that might tend to change the accretion rate.
First, the inner disk might be inclined differently from the masing disk.
If $i = 45^\circ$, then the accretion rate can be pushed as low as $0.0012
\msun \yr^{-1}$; the order of magnitude change in $\dot{M}$ again follows
from the nonlinear dependence of $\dm$ on $\cos(i)$ in eq.(\ref{FNEQN}).
It is also possible, however, that the inclination of the inner disk is
larger than $82^{\circ}$.  Second, the obscuring dust grains might well have
a different size distribution than in the local interstellar medium.
The sense of the effect of grain size evolution, however, at least in
the grain models of \cite{ld93}, is to decrease $A_H/A_K$ and so increase
the required mass accretion rate.

Third, some of the near-infrared emission may be reprocessed light from
the central engine, either in a disk or in a wind (e.g. \cite{kk94}).
Crudely speaking, the light must be reprocessed at $r \sim r_{1500}
\sim 2.8 \times 10^3 L_{42}^{1/2} G M/c^2$ corresponding to a
characteristic temperature of $1500 K$ for near IR emission (dust
grains much smaller than $1\micron$ can produce H and K band emission
from a somewhat larger radius because they radiate inefficiently in the
infrared).  When is reprocessing important in a disk?  Suppose
$\mu(r,\phi)$ is the ``obliquity'' of the disk, which is to say, the
angle between the normal to the disk surface and the radius vector, $f$
is the fraction of intercepted radiation that is thermalized in the
disk (so $1 - f$ is an effective albedo).  Then reprocessing is
dominant if $r \gtrsim (G M/c^2)(\epsilon f \cos\mu)^{-1}$.  For a
thin, flared disk $\cos\mu \sim H/R$.  One can then show that, in a
standard $\alpha$ disk model over a large range of $\dm$, reprocessing
is not dominant at $r \sim r_{1500}$.  

If the disk is warped then $\cos\mu$ can be larger and reprocessing can
dominate internal heating.  It is not possible to survey the warp
parameter space, since a warp is described by two free functions: the
inclination $i(r)$ and position angle $p(r)$.  A broad range of
behavior is possible.  For example, the warp could be made sufficiently
large so as to cover the central engine (raising the inferred accretion
rate); it could also be arranged so that the inner disk is nearly face
on (lowering the inferred accretion rate).

As an illustration, however, we have constructed a ``naive'' warped
disk model with only an inclination warp ($p(r) = {\rm const}.$) in which the
reprocessed light is reradiated as a black body.  The disk extends from
$r = 6 G M/c^2$ to $0.26 \pc$, as does the flat disk model, and has
constant $\dm$.  The effective temperature $T_e$ of the disk is given
by $T_e^4 = T_{int}^4 + f_{obl} T_{ext}^4$, where $T_{int}$ is the effective
temperature in the absence of external heating, and
\begin{equation}
\sigma T_{ext}^4 = {\eps \dm c^2\over{4\pi r^2}}
\end{equation}
and $\eps = 0.1$.  Here $f_{obl}$ is the obliquity factor.
If the disk is illuminated on the side facing us, then $f_{obl} \simeq
\sin(\phi) |d i/d\ln r|$, where $\phi$ is the disk orbital phase.  If
the disk is illuminated on the side facing away from us, as it is for
half its orbit, then the reprocessing only adds a flux $\sigma
T_{ext}^4/\tau$ to our side of the disk, and so, assuming that $\tau \gg
(T_{ext}/T_{int})^4$, we set $f_{obl} = 0$.

Figure 2 shows the resulting thin, warped disk spectrum.  The warp
extends from $r = 100 G M/c^2$ to $r = 0.26\pc$ with $d i/d\ln r =
0.05$.  The accretion rate is $8 \times 10^{-4}\msun\yr^{-1}$.  Three
effects combine here to reduce the required accretion rate: (1)
reprocessing enhances the $2 \micron$ flux; (2) reprocessing make the
spectrum shallower at $2 \micron$ thus reducing the reddening
correction; (3) the inclination warp changes the inclination of the
inner disk from $82^{\circ}$ to $61^{\circ}$, thus making the reprocessing portion
of the disk more face-on.

Our warped disk model is naive in the sense that the reprocessed
emission does not emerge as a blackbody.  In particular, dust grains in
the disk atmosphere can cause half of the flux intercepted by the disk
to reemerge as a dilute blackbody with $T > T_e$.  This effect has been
considered in the context of quasar disks by \cite{s89} and \cite{esp89};
the same effect has been considered in the context of circumstellar disks
by \cite{cg97}.  In Phinney's quasar models the resulting spectrum has
a notch at $\sim 1\micron$ related to the onset of dust sublimation at
$T \approx 1800\K$.  Since the spectrum might then be declining at $H$
and $K$ bands, this could reduce the reddening correction and thus also
reduce the required accretion rate.

Reprocessing is ultimately limited, however, by the luminosity of the
central accretion flow.  Figure 1 also shows a thin disk spectrum for a
face-on disk truncated at $r = 6 \M$, with $\dm = 7 \times 10^{-6} \msun
\yr^{-1}$, corresponding to the model of \cite{nm95} with $\alpha = 0.1$.
It would be difficult to produce the observed infrared flux from such
a low accretion rate disk even with a highly optimized arrangement of
reprocessing material.

A reprocessing model for the H and K band flux can be tested via long
term monitoring.  If the light is reprocessed, then IR variability
should be highly correlated with X-ray variability, with delays on
order the light travel time $r_{1500}/c \simeq 5.5 \times 10^5
L_{42}^{1/2} \sec$.

Finally, while we have not explicitly fit the X-ray emission using a
thin disk plus corona model (e.g. \cite{hm91}), it is generally
acknowledged that it is possible to do so, since there are a number of
free parameters associated with the corona.

\subsection{Advection-Dominated Flow Model}

An alternative to the thin disk plus corona model for the inner
accretion flow is the two-temperature advection-dominated accretion
flow model (ADAF; \cite{i77,ny94,ny95b,acklr95}; see \cite{nar97a,nmq98} 
for reviews).  ADAFs are optically thin and geometrically thick, with
$H/R \sim 1$.  They have been used to successfully fit the spectral
energy distribution of accretion flows onto both stellar mass and
supermassive black holes.  In their simplest form, ADAFs approximate
the flow as non-radiative, with nested spherical surfaces moving
radially inwards with an angular velocity given by the value on the
equatorial plane. Radial pressure gradients combine with centrifugal
force to oppose gravity and radial motion is fast, though subsonic
until it approaches the event horizon.  There is hypothesized to be a
large viscous stress ($\sim\alpha P$, where $P$ is the pressure) that
transports angular momentum radially outward.

The ADAF approach has some dynamical shortcomings.  Firstly, it does
not take account of angular velocity gradients on spherical surfaces.
This appears to be a reasonable approximation far from the black hole
in the self-similar region of the flow (cf. \cite{ny95a}), but the
approximation may break down close to the hole, especially close to
the symmetry axis.  This is relevant for understanding whether the
flow forms a funnel and a jet.  The indications are that there is no
funnel when $\alpha\gtrsim 0.01$ (\cite{nar97b,gp98,pg98}, though this
conclusion is not certain). Secondly, the solutions assume vertical
hydrostatic equilibrium, which is suspect in the supersonic region close
to the hole.  This approximation should not have direct impact on the
emergent radiation, though.  Finally, in these solutions, the gas has
positive Bernoulli constant and always has enough energy to escape
the hole (see \cite{ny94,ny95a,nar97b}).  This arises because of the
inevitable, outward radial transport of energy by the viscous stress,
which is balanced exactly by inward advective transport of energy.
This balance must be maintained everywhere.  If it is not, then it is
possible that energy may be carried away by a wind (see \cite{bland98},
and in preparation).

%If it turns out that
%this energy has no sink at $r_{tr}$, the outer edge of the ADAF where
%the disk changes from thick to thin, then it is possible that the energy
%actually drives a wind (see \cite{bland98}, and in preparation).

%%I added this para to balance the previous one --- RN
Counterbalancing these shortcomings is the fact that the ADAF is the
only model of hot accretion flows that attempts to calculate the
dynamics, thermal structure and radiative spectrum of the gas
self-consistently.  Other approaches, e.g. coronal models, leave most
details of the dynamics unspecified and allow far greater freedom in
choosing the properties of the radiating plasma.

\cite{la96} proposed a model for \ngc which comprises a thin disk
extending outward from the transition radius $r_{tr}$ and an ADAF
extending inward to the black hole.  The mass accretion rate in the
model is essentially fixed by the X-ray data point to be of order $0.014
\am1 \msun \yr^{-1}$; Lasota et al. take $\am1 = 1$.  One free parameter
remains: $r_{tr}$.  At the time \cite{la96} was written the data did not
constrain $r_{tr}$.  Two results have since appeared that do: the $22
\ghz$ upper limit of \cite{h98}, and the IR detections of \cite{cb97}.
There has also been considerable improvement in the calculation of
the ADAF model spectral energy distribution.  Relativistic effects
(\cite{gp98}, and references therein) and advective heating of the
electrons (\cite{nak97}) are now included; details of the computation are
described in \cite{nar98}, which applies the ADAF model to the Galactic
Center source Sgr A$^*$.

Figure 3 shows the spectral energy distribution of \ngc corresponding to
three ADAF models, with $r_{tr} = (10, 10^{1.5}, 10^2) \times
2 G M/c^2$ and $\dm = (0.012, 0.008, 0.006) \msun \yr^{-1}$, which we
call models $1,1.5,$ and $2$.  The disk outside $r_{tr}$ is thin, flat,
and radiates like a blackbody.  All models use the standard values for
other parameters: $\alpha = 0.3, \beta = 0.5,$ and $\delta = 10^{-3}$
($\beta \equiv p_{gas}/p_{tot}$, $\delta \equiv$ fraction of dissipation
going into electrons).  Model $1$ has the most (least) flux in the
optical (radio).  Evidently model $1$ fits the IR data best and does
not violate the $22 \ghz$ upper limit, although the model $1.5$ also
gives an acceptable fit to the IR data and may have better agreement
with the spectral index of the scattered optical light.  Model 2 is
close to violating the $22 \ghz$ upper limit.

The ADAF model predicts the flux from the central accretion flow
at several wavelengths that may be observed in the near future.
At $10 \micron$ the observed flux should not be below the thin disk
value (see equation [\ref{FNEQN}]) of $(200, 150, 130) \cos(i) \mjy$
for models $(1,1.5,2)$; it can be larger because of reradiation by
dust.  At $22 \ghz$ the flux should be $(0.017, 0.082, 0.23) \mjy$
with $s \equiv d\ln f_\nu/d\ln\nu \approx 1.5$ for all three models.
Higher frequency interferometry may be possible; at $100 \ghz, f_\nu =
(0.13, 0.65, 1.5) \mjy$, with $s = (1.8, 1.4, 1.3)$, and at $690 \ghz,
f_\nu = (3.4, 10., 18.) \mjy, s = (1.6, 1.3, 1.0)$.  At $5500 \AA$ the
expected flux is $(38,8.4,0.59) \mjy$ with $s = (-0.39, -2.0, -2.2)$;
recall that \cite{w95} find $s = -1.1$.  The model is adjusted so that
the flux at $10 \kev$ fits the observations; at $10 \kev$ the predicted
photon index is $(1.9,1.9,2.0)$, which may be compared with the observed
index of $1.78 \pm 0.29$.  Finally, the expected flux at $30 \kev$ is
$(3.1, 3.4, 3.4) \times 10^{-6} {\rm ph} \cm^{-2} \sec^{-1} \kev^{-1}$,
and the photon index for all three models is $2.0$.

\subsection{Ion Torus Model}

A related, though distinct, alternative to the ADAF is the ion torus
(\cite{rees82}).  As in the ADAF, it is assumed that the plasma is
two-temperature with most of the pressure being supplied by the ions
(see \cite{shap76} for the first application of the two-temperature idea
to accretion) and that the radiative losses are small.  The viscosity
is assumed to be small enough (low $\alpha$), however, that poloidal
motion can be ignored in the equation of hydrostatic equilibrium. In
its simplest manifestation, the specific angular momentum is constant
and the surface of the torus is a surface of constant binding energy. As
the radiative losses are small, the binding energy is quite low (relative
to $c^2$ per unit mass) and plasma flows onto the black hole through an
equatorial cusp with this energy and consistent with having lost little
of its binding energy.  This approach can accommodate the peculiarly
relativistic features of the Kerr metric (\cite{ajs78}).

Following previous work on thick disks, the ion torus model simply
specifies the angular momentum profile rather than solving an angular
momentum equation, and this is a serious weakness.  Also, the model has no
ready prescription for describing the gas inflow and, consequently,
the thermal structure.  In addition, tori of this sort are subject to
hydrodynamical instability (\cite{pap84}), which calls into question
the assumption of slow angular momentum transport.

Nonetheless, ion tori do have one feature that is important in the present
context: they naturally create a pair of funnels in which magnetic flux
can be concentrated so as to extract energy from the central spinning
black hole and power the jets.  According to the work of \cite{nar97b}
and \cite{gp98}, funnels are likely to be most prominent in low-$\alpha$
flows (ion tori), and are much less pronounced in high-$\alpha$ flows
(standard ADAFs).

\subsection{General Constraints}

The failure to detect radio emission from the location of the black
hole has some more general implications.  Although we have demonstrated
that ADAF models can be made to be consistent with this upper limit,
alternative models that admit a non-thermal
distribution of relativistic electrons to be accelerated directly are quite
tightly constrained.

Let us suppose that there is a radio photosphere where the optical depth
to synchrotron self-absorption becomes unity. Let it be located at a
radial distance $r_p=10^{14}r_{p14}$~cm from the black hole.  The 22~GHZ
upper bound on the flux implies that the radio brightness temperature
of this photosphere must satisfy
\begin{equation}
T<2\times10^{11}r_{p14}^{-2}{\rm ~K}
\end{equation}
(\cite{he97}).  The corresponding energy of the electrons emitting at
$22 \ghz$ is $\sim 3 k T \sim E_p\sim70B_p^{-1/2} ~MeV$.  These
electrons radiate at $\sim \gamma^2 B \mhz$ (cf. \cite{rl79}), and so
we conclude that the magnetic field strength at the photosphere
satisfies $B_p=B(r_p)\gtrsim r_{p14}^4$~G, (assuming that it is
significantly less than the cyclotron field, $\sim7000$~G).  In other
words the magnetic pressure must satisfy $P_{{\rm m p}}\gtrsim
0.04r_{p14}^8$, measured in dyne cm$^{-2}$.  We use the theory of
synchrotron radiation to solve for the radius of the 22~GHz
photosphere:
\begin{equation}
r_{p14}\sim10f_eP_{{\rm m p}}^2,
\end{equation}
where $f_e$ is the ratio of the emitting electrons to the magnetic
pressure.  Now, in order not to exceed the radio flux upper limit, we
require that $r_{p14}\lesssim1.3f_e^{0.07}\sim20(GM/c^2)f_e^{0.07}$
and, equivalently, $P_{{\rm m p}}\lesssim1.1 f_e^{-0.03}$~dyne
cm$^{-2}$.  As emphasized below, however, typical pressures in accretion flows
are generally expected to exceed $\sim1$~dyne cm$^{-2}$ by a large
factor, so this is a fairly stringent constraint.

ADAF models satisfy this constraint because they ignore extraneous
relativistic particle acceleration so that the electron distribution
function is thermal and exponentially declining at energies below $E_p$.
(\cite{mq97} have shown that the thermalization time scale for the
electrons in an ADAF is shorter than the flow time scale, so that the
assumption of a thermal particle distribution is consistent.)  What we see
from the analysis in this section is that this assumption is necessary.

Disk corona models generally do postulate relativistic particle
acceleration.  The accelerated particles might be expected to cause
observable radio emission, and so the limit on the 22 GHz brightness
temperature is a serious constraint.

There are some processes which can suppress radio synchrotron emission.
The Razin effect is important at densities above $\sim10^9B_p$~cm$^{-3}$.
Secondly, relativistic Coulomb interactions and synchrotron loss may
cool freshly accelerated electrons to keep their steady pressure at
a low level.  This happens on time scales competitive with the inflow
timescale. Thirdly, extrinsic free-free absorption may extinguish
the radio source.  Two difficulties with this explanation are that
the southern jet can be seen through the accreted gas at a projected
perpendicular distance from the black hole of $\sim6000 G M/c^2$ and
that there is no observed soft X-ray absorption.

\section{Jet}

It is clearly of interest to relate the observed VLBI jets to the gas
flow close to the black hole. If we consider the northern VLBI component,
we find that, at a projected distance of $r_\perp\sim4\times10^{16}$~cm,
the 22~GHz flux density is $\sim3$~mJy. If we suppose that the unresolved
source is the base of a jet with an opening angle of $\sim10^\circ$ (cf.
\cite{he97})propagating nearly perpendicular to the line of sight with
a bulk Lorentz factor $\Gamma$, then the radio brightness temperature
is only $\sim2\times10^9$~K.  Even allowing for relativistic beaming
this is too low for synchrotron self-absorption to be important and
it seems reasonable to assume that both radio components are optically
thin at this frequency. Let us now assume that the relativistic energy
density in the jet is not much smaller than the magnetic energy density.
A straightforward calculation relates the minimum power carried by the
relativistic electrons (and, quite possibly, positrons) in the jet to
$\Gamma$ and to the ratio of the electron energy density to
the magnetic energy density $\beta_e$ (assumed $>1$):
\begin{equation}
L_{{\rm jet}}\sim7\times10^{42}\beta_e^{0.4}\left({\Gamma\over10}\right)^{3.4}
{\rm erg~s}^{-1}
\end{equation}
(see, e.g., \cite{rdbsaasfee}).
The strong dependence of the minimum jet power upon the Lorentz factor
is due to a combination of the bulk energy and the diminution of the
observed flux by relativistic beaming when the observer is located
roughly perpendicular to the jet direction.  We see that if jets have
ultrarelativistic flow speeds, then they probably carry off significantly
more power than is radiated directly.

The power in the jet may be extracted directly from the spin of the
black hole (\cite{bz77}).
If instead, the power is derived from the
accretion flow, then the jet implies a certain minimum mass accretion rate
\begin{equation}\label{JETMDOT}
\dot M\sim1.2\times10^{-3}\ep1^{-1}\beta_e^{0.4}
\left({\Gamma\over10}\right)^{3.4} ~M_\odot\,{\rm yr^{-1}}.
\end{equation}
Notice that this is similar to the estimates of $\dot M$ obtained earlier
by fitting the observed infrared spectrum.

The pressure in the jet
$\sim0.07\beta_e^{0.4}(\Gamma/10)^{1.4}(r/4\times10^{16}{\rm
cm})^{-2}$~dyne cm$^{-2}$ must be confined transversely. If we
evaluate this at the transition radius of the ADAF solution, then the
jet pressure is $\sim700(\Gamma/10)^{1.4}\beta_e^{0.4}$~dyne
cm$^{-2}$. For comparison, the pressure in the ADAF flow is
\begin{equation}
P\sim 8000\left({\dot M\over10^{-2}{\rm M}_\odot{\rm yr}^{-1}}\right)\am1^{-1}\left(
{r_{\rm tr}\over60GM/c^2}\right)^{-5/2}{\rm dyne~cm}^{-2} .
\end{equation}
We note that, although the formation of funnels appears to be problematic
in ADAFs with $\alpha\gtrsim 0.01$ (\cite{nar97b,gp98}), the pressure
required to confine a jet able to account for the observed radio
sources is probably present.  An ion torus, by contrast, naturally
forms a torus and can exert more pressure than a high-$\alpha$
ADAF as long as it is unable to cool.  If the accretion rate is
$\lesssim10^{-4}$~M$_\odot$~yr$^{-1}$, however, it does not seem possible
to supply enough pressure to confine a relativistic jet.

\section{Discussion}

We have argued that the outer, masing disk in \ngc is unlikely to
be in a steady state because of its long viscous timescale.  Thus it
is questionable whether conditions in the masing disk can be used to
estimate the mass accretion rate onto the central black hole.

We have fit several models to the spectral energy distribution of the
nucleus in order to better constrain the accretion rate.  The ADAF
model is able to fit all the data as long as $\dm \approx 10^{-2}
\msun \yr^{-1}$, and the transition radius (where the accretion flow
makes the transition from a thin disk to an ADAF) is $\approx (10-100)
\M$.  The ADAF accretion rate is basically fixed by the X-ray
luminosity, but it is nicely consistent with the observed infrared
flux as long as the infrared is generated by a thin, flat disk with
inclination close to that of the masing disk.

We have also fit pure thin disk models to the infrared data.  (The
X-ray flux would presumably be generated by a hot corona, but we have
not constructed an explicit corona model, since coronae are not as
tightly constrained as ADAFs.)  The inferred accretion rate is strongly
inclination dependent.  For $i=82^o$, $\dot M=0.013 ~M_\odot\,{\rm
yr^{-1}}$.  A lower accretion rate of $0.0012 \msun \yr^{-1}$ is obtained
by giving the disk an inclination of $45 ^{\circ}$, although such a low
inclination seems unlikely given the orientation of the large scale jet.
The accretion rate can be further lowered by giving the disk a warp
that reprocesses radiation from the central source into the infrared.
Since the warp parameter space is large, we have calculated the spectrum
of a representative model disk with a modest warp, and the implied
accretion rate is $8 \times 10^{-4} \msun \yr^{-1}$.  The accretion
rate might be pushed lower by a somewhat artificial positioning of
reprocessing gas at the right radius; this could push the accretion rate
as low as $7 \times 10^{-5} (2.1)^s \ep1^{-1} \msun \yr^{-1}$, where $s$
is the slope of the intrinsic spectrum.  For $s = -1$, $\ep1 = 1$, and
$\am1 = 1$ (larger values seem unlikely except possibly in the event of
gravitational instability), this minimal accretion rate is still larger
than the $7 \times 10^{-6} \am1 \msun \yr^{-1}$ obtained by \cite{nm95}
from consideration of the masing disk.

The minimum power in the jet in \ngc also constrains the mass accretion
rate, provided the jet power is derived from the accretion flow and not
from the spin of the black hole.  The resulting estimate for $\dot M$
(cf. equation [\ref{JETMDOT}]) is similar to that obtained by fitting
the infrared spectrum.

The limit on the 22 GHz radio flux in \ngc constrains the transition
radius in any ADAF to less than 100 Schwarzschild radii, as emphasized
by \cite{h98}.  As we show in this paper, the flux limit
also constrains corona models.  If the observed X-ray emission in \ngc
is from a corona, then the corona must be primarily thermal rather
than non-thermal, since otherwise it would produce more radio flux
than observed.  This limits the kinds of relativistic acceleration one
can invoke in the corona.

New data at several different frequencies would give interesting
information on the nature of the central accretion flow in NGC 4258.
The ADAF model predicts a complete spectrum whose amplitude and slope can
be compared with new data in the radio, millimeter, infrared, and X-ray;
the predicted values are given in \S 4.

It would be particularly interesting to know if the near infrared flux
is reprocessed or not.  This can be tested by looking for variation on
a timescale $t_r = r_{1500}/c \simeq 5.5 \times 10^5 L_{42}^{1/2}
\sec$, which is about a week.  Since the viscous timescale-- indeed,
even the dynamical timescale-- is much longer than this in the disk at
$r_{1500}$, we would not expect significant variation in this component
of the flux unless the infrared flux is reprocessed, or the infrared
flux is produced nonthermally.  If a significant fraction of the H and
K band flux is reprocessed, this would rule out our simplest models and
we would be forced to invoke a warped disk, wind, or other arrangement
of reprocessing gas.  Higher angular resolution mid-infrared data would
also be interesting, since a flat disk predicts a steeper spectrum in
the mid-infrared than a warped, reprocessing disk.

Finally VLBI observations at lower frequencies than $22 \ghz$ may tighten
the general constraints on relativistic particle acceleration close to
the black hole and exclude free-free absorption as a factor.  They may
also be able to determine the onset of self-absorption in the jet and
thence provide a stronger constraint on the pressure and jet power by
allowing us to solve for the magnetic field strength. Further monitoring
of the variability in both the northern and the southern component may
also help determine the jet speed and elucidate the relationship between
the jets and gas flow within $\sim100$ gravitational radii of the central
black hole.

\acknowledgments 

This work was supported in part by NSF grants AST
94-23209, AST 95-29170, and NASA grants NAG 5-2837, NAG5-7007.  We thank
A. Esin for assistance with the ASCA data.

\clearpage

\begin{figure}
\plotone{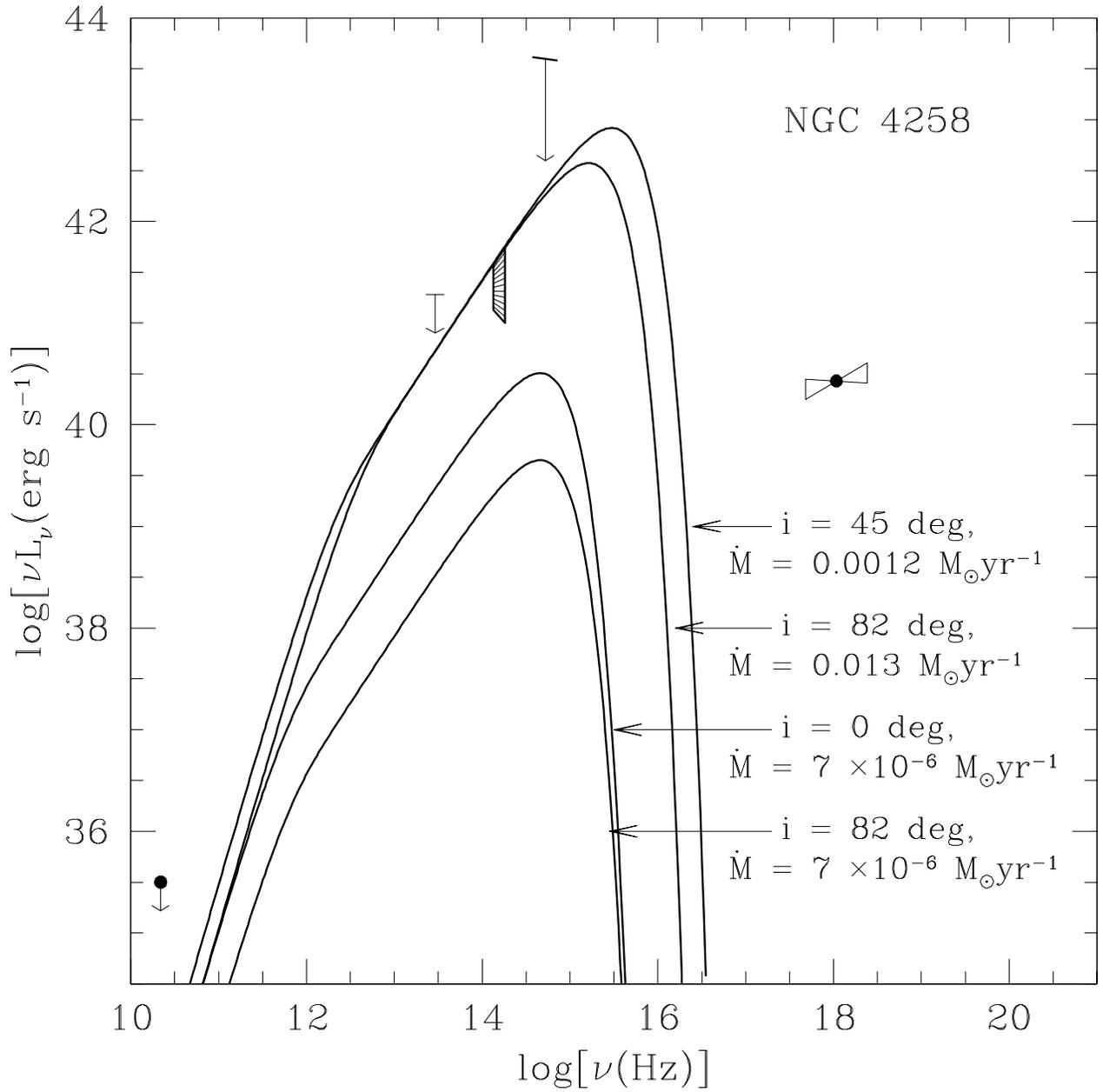}
\caption{
The spectral energy distribution of the nucleus of \ngc.  The data
points are described in the text (see \S 2).  The models are 
flat, thin disks with inclination $i = (45,82,0) ^{\circ}$ and
$\dm = (0.0012, 0.013, 7 \times 10^{-6}) \msun \yr^{-1}$.
}
\end{figure}

\begin{figure}
\plotone{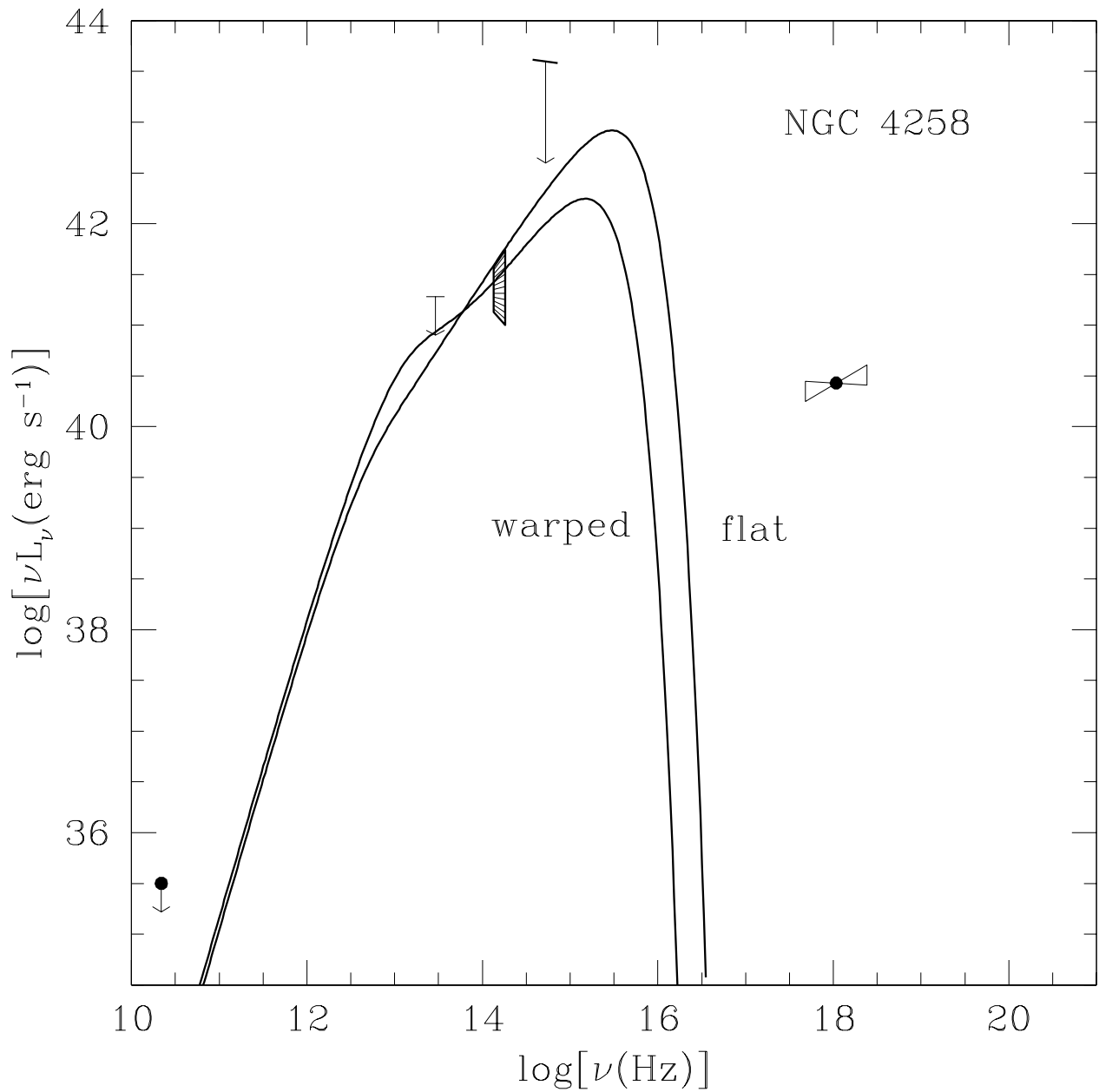}
\caption{
Same as Figure 1, but for a warped, reprocessing disk with $\dm =
0.0008 \msun \yr^{-1}$ as well as the flat disk model with 
$i = 82 ^{\circ}$ and $\dm = 0.013 \msun \yr^{-1}$.
}
\end{figure}

\begin{figure}
\plotone{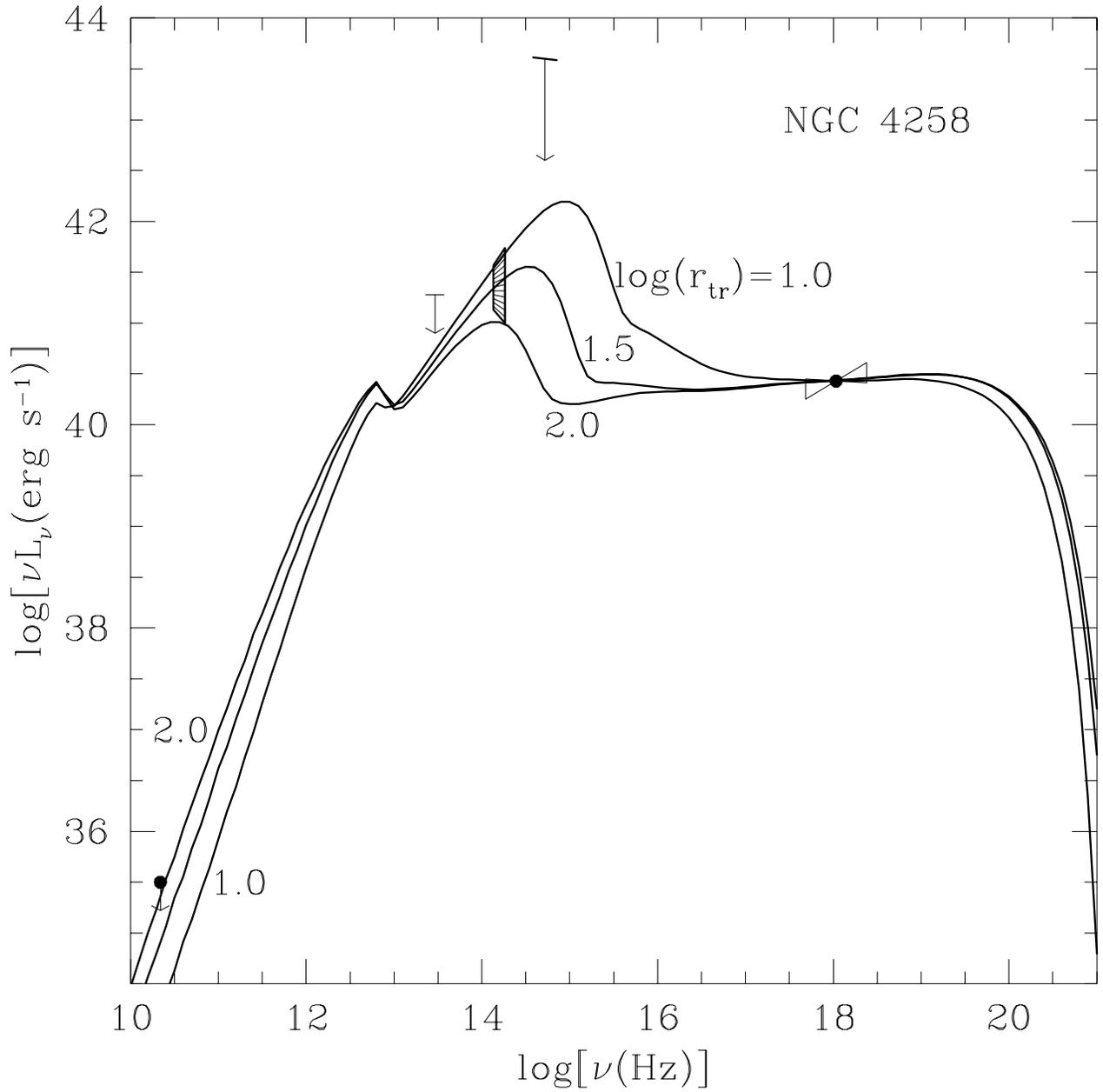}
\caption{
The models are ADAFs fit
with $r_{tr} = (10, 10^{1.5}, 10^2) 2 G M/c^2$ and $\dm = (0.012,
0.008, 0.006) \msun \yr^{-1}$.  As the transition radius
moves in the optical/IR flux increases.
}
\end{figure}


\begin{thebibliography}{}

\bibitem[Abramowicz, Jaroszy\'nski \& Sikora (1978)]{ajs78} 
	Abramowicz, M. Jaroszy\'nski, M. \& Sikora, M. 1978, \aap, 63, 221

\bibitem[Abramowicz (1995)]{acklr95} 
	Abramowicz, M., et al. 1995, \apj, 438, L37.

\bibitem[van Albada \& van der Hulst (1982)]{vv82} van Albada, G D., \& van der
	Hulst, J. M. 1982, \aap, 115, 263

\bibitem[Blandford (1990)]{rdbsaasfee} Blandford, R.D. 1990, in Active Galactic
	Nuclei, Saas-Fee Advanced Course 20. (Berling: Springer-Verlag).

\bibitem[Blandford (1998)]{bland98} Blandford, R. D. 1998 Proc. Maryland Conference
      on Astrophysics. ``Some Like it Hot''. ed. S. Holt \& T. Kallman, New York: AIP
 
\bibitem[Blandford \& Znajek (1977)]{bz77} Blandford, R.D., \& Znajek, R. L.
	1977, MNRAS, 179, 433.

\bibitem[Cao \& Jiang (1997)]{cj97} Cao, X., \& Jiang, D.R. 1997,
	\aap, 322, 49

\bibitem[Cardelli et al. (1989)]{ccm89} Cardelli, J.A., Clayton, G.C.,
	\& Mathis, J.S. 1989, \apj, 345, 245

\bibitem[Cecil et al. (1995a)]{cwd95} Cecil, G., Wilson, A.S., \&
	DePree, C. 1995, \apj 440, 181

\bibitem[Cecil et al. (1992)]{cwt92} Cecil, G., Wilson, A.S., \& Tully,
	R.B., 1992, \apj, 390, 365

\bibitem[Cecil et al. (1995b)]{cmv95} Cecil, G., Morse, J.A., \& Veilleux,
	S. 1995, \apj, 452, 613

\bibitem[Chary \& Becklin (1997)]{cb97} Chary, R., \& Becklin, E. E.
	1997, ApJ Lett, 485, L75

\bibitem[Chiang \& Goldreich (1997)]{cg97} Chiang, E.I., \& Goldreich, P.
	1997, \apj, 490, 368

\bibitem[Cizdziel et al. (1985)]{cwb85} Cizdziel, P.J., Wynn-Williams,
	C.G., \& Becklin, E.E. 1985, \aj, 90, 731

\bibitem[Claussen et al. (1984)]{chl84} Claussen, M.J., Heiligman,
	G.M., \& Lo, K.Y., Nature, 310, 298

\bibitem[Claussen \& Lo (1986)]{cl86} Claussen, M.J., \& Lo, K.Y., 
	\apj 308, 592

\bibitem[Court\`es \& Cruvellier (1961)]{cc61} Court\`es, G., \& Cruvellier,
	P. 1961, CR Acad. Sci. Paris, 253, 218

\bibitem[Gammie \& Popham (1998)]{gp98} Gammie, C.F., \& Popham, R. 1998,
	\apj, 498, 313.

\bibitem[Greenhill et al. (1995c)]{gr95c} Greenhill, L.J. et al., 1994,
	in ``Highlights of Astronomy, Vol. 10'' (Dordrecht: Kluwer)
	p. 531

\bibitem[Greenhill et al. (1995a)]{gr95a} Greenhill, L.J. et al., 1995,
	\apj, 440, 619

\bibitem[Greenhill et al. (1995b)]{gr95b} Greenhill, L.J. et al., 1995,
	\aap, 304, 21

\bibitem[Haardt \& Maraschi (1991)]{hm91} Haardt, F., \& Maraschi, L.
	1991, \apj, 380, 51

\bibitem[Haschick \& Baan (1990)]{hb90} Haschick, A.D., \& Baan, W.A.
	1990, \apj, 355, L23

\bibitem[Haschick et al. (1994)]{ha94} Haschick, A.D., Baan, W.A., \&
	Peng, E.W. 1994, \apj, 356, 149

\bibitem[Herrnstein (1997)]{h97} Herrnstein, J.R. 1997, PhD thesis,
	Harvard University

\bibitem[Herrnstein et al. (1998)]{h98} Herrnstein, J.R. et al. 1998,
	\apj, 497, 69

\bibitem[Herrnstein, Greenhill, \& Moran  (1996)]{hgm96} Herrnstein, 
	J.R., Greenhill, L., \& Moran, J. 1996, \apj, 468, L17

\bibitem[Herrnstein et al. (1997a)]{he97} Herrnstein, J.R., et al.,
	1997,  ApJ Lett, 475, L17

\bibitem[Herrnstein et al. (1997c)]{hp97} Herrnstein, J.R., et al. 1997, 
	abstract published in BAAS, 191, \#25.07	

\bibitem[Ichimaru (1977)]{i77} Ichimaru, S. 1977 \apj, 214, 840.

\bibitem[K\"onigl \& Kartje (1994)]{kk94} Konigl, A., \& Kartje, J.F.,	
	1994, \apj, 434, 446

\bibitem[van der Kruit et al. (1972)]{kom72} van der Kruit, P. C.,
	Oort, J.H., \& Mathewson, D. S. 1972, \aap, 21, 169

\bibitem[Kumar (1997)]{pk97} Kumar, P. 1997, \apj, submitted (astro-ph
	9706063)

\bibitem[Laor \& Draine (1993)]{ld93} Laor, A., \& Draine, B.T.,
	1993, \apj, 402, 441

\bibitem[Lasota et al. (1996)]{la96} Lasota, J.P., et al. 1996,
	\apjl, 462, 142

\bibitem[Mahadevan \& Quataert (1997)]{mq97} Mahadevan, R., \&
	Quataert, E. 1997, \apj, 490, 605

\bibitem[Makishima et al. (1994)]{ma94} Makishima, K., et al., 1994,
	PASJ, 46, L77

\bibitem[Maoz \& McKee (1997)]{mm97} Maoz, E., \& McKee, C. F. 1997,
	\apj, submitted (astro-ph 9704050)

\bibitem[Martin et al. (1989)]{mrnl89} Martin, P., Roy, J.-R.,
	Noreau, L., \& Lo, K.Y. 1989, \apj, 345, 707

\bibitem[Miyoshi et al. (1995)]{mi95} Miyoshi, M., et al. 1995,
	Nature, 373, 127

\bibitem[Moran et al. (1995)]{mo95} Moran, J.M., et al. 1995,
	PNAS, 92, 11427

\bibitem[Nakai et al. (1993)]{nim93} Nakai, N., Inoue, M., \& 
	Miyoshi, M., 1993, Nature, 361, 45

\bibitem[Nakamura et al. (1997)]{nak97} Nakamura, K.E. et al. 1997,
	\pasj, 49, 503.

\bibitem[Narayan (1997)]{nar97a} Narayan, R. 1997, in Proceedings of IAU
	Colloquium 163, Accretion Phenomena and Related Objects,
	A. S. P. Conference Series, eds. D. T. Wickramasinghe,
	L. Ferrario, \& G. V. Bicknell, p.75

\bibitem[Narayan \& Yi (1994)]{ny94} Narayan, R., \& Yi, I. 1994,
	\apj, 428, 13.

\bibitem[Narayan, Kato \& Honma (1997)]{nar97b} Narayan, R., Kato, S. \& Honma, F.,
	1997, \apj, 476, 49

\bibitem[Narayan \& Yi (1995a)]{ny95a} Narayan, R., \& Yi, I. 1995,
	\apj, 444, 231.

\bibitem[Narayan \& Yi (1995b)]{ny95b} Narayan, R., \& Yi, I. 1995,
	Nature, 374, 623.

\bibitem[Narayan et al. (1998)]{nar98} Narayan, R., et al. 1998, 
	\apj, 492, 554

\bibitem[Narayan et al. (1998)]{nmq98} Narayan, R., Mahadevan, R., \&
	Quataert, E. 1998, to appear in ``The Theory of Black Hole
	Accretion Discs'', eds. M.A.Abramowicz, G.Bjornsson, \&
	J.E.Pringle (Cambridge) (astro-ph 9803141)

\bibitem[Neufeld \& Maloney (1995)]{nm95} Neufeld, D.A., \& Maloney,
	P.R., 1995, ApJ Lett, 447, L17

\bibitem[Papaloizou \& Pringle (1984)]{pap84} Papaloizou, J. C. B. \& 
      Pringle, J. E., 1984, \mnras, 208, 721

\bibitem[Phinney (1989)]{esp89} Phinney, E. S. 1989, in
	Theory of Accretion Disks, eds. W Duschl et al.
	(Dordrecht: Kluwer), p. 457

\bibitem[Popham \& Gammie (1998)]{pg98} Popham, R., \& Gammie, C.F.
	1998, \apj, in press (astro-ph 9802321).

\bibitem[Pringle (1981)]{jep81} Pringle, J. 1981, \araa, 19, 137

\bibitem[Rees et al. (1982)]{rees82} Rees, M. J., Begelman, M. C., Blandford, R. D.
      \& Phinney, E. S. 1982, Nature 295 17

\bibitem[Rieke \& Lebosky (1978)]{rl78} Rieke, G. H., \& Lebofsky,
	M.J., 1978, \apj, 220, L37

\bibitem[Rybicki \& Lightman (1979)]{rl79} Rybicki, G. B., \& Lightman, A. P.,
	Radiative Processes in Astrophysics (New York: Wiley).

\bibitem[Sanders et al. (1989)]{s89} Sanders, D.B., et al. 1989,
	\apj, 347, 29

\bibitem[Shakura \& Sunyaev (1973)]{ss73} Shakura, N.I., \& Sunyaev,
	R. A. 1973, \aap, 24, 337

\bibitem[Shapiro et al 1976]{shap76} Shapiro, S. L., Lightman, A. P. \& Eardley, D. M.
      1976 \apj, 204, 187

\bibitem[Spitzer (1978)]{spitzer} Spitzer, L., 1978, Physical Processes
	in the Interstellar Medium (New York: Wiley)

\bibitem[Watson \& Wallin (1994)]{ww94} Watson, W.D., \& Wallin, B.K.,
	1994, ApJ Lett, 432, L35

\bibitem[Wilkes et al. (1995)]{w95} Wilkes, B.J., et al. 1995, \apj,
	455, L13

\end{thebibliography}
\end{document}